\documentclass[12pt]{article}
\usepackage{graphicx}
\usepackage{xspace}
\usepackage{amsmath}

\newcommand\pubnumber{DPF2015-92}
\newcommand\pubdate{\today}

\def\napoli{Northeastern University, Department of Physics\\
360 Huntington Avenue, Boston, MA 02115, USA}
\def\support{\footnote{Work supported by the US Department of Energy.}}

\def\Title#1{\begin{center} {\Large #1 } \end{center}}
\def\Author#1{\begin{center}{ \sc #1} \end{center}}
\def\Address#1{\begin{center}{ \it #1} \end{center}}

\newcommand\pubblock{\rightline{\begin{tabular}{l} \pubnumber\\
         \pubdate  \end{tabular}}}
\newenvironment{Abstract}{\begin{quotation}  }{\end{quotation}}
\newenvironment{Presented}{\begin{quotation} \begin{center} 
             PRESENTED AT\end{center}\bigskip 
      \begin{center}\begin{large}}{\end{large}\end{center} \end{quotation}}

\def\beq{\begin{equation}}
\def\eeq#1{\label{#1}\end{equation}}
\def\eeqn{\end{equation}}

\def\beqa{\begin{eqnarray}}
\def\eeqa#1{\label{#1}\end{eqnarray}}
\def\eeqan{\end{eqnarray}}

\let\bar=\overbar

\def\Dslash{\not{\hbox{\kern-4pt $D$}}}
\def\dslash{\not{\hbox{\kern-2pt $\del$}}}

\def\msb{{\bar{\ssstyle M \kern -1pt S}}}

\input pennames-ital.sty
\input ptdr-definitions.sty
\newcommand{\vmet}{\ensuremath{\vec{E}_\mathrm{T}}^{\text{miss}}\xspace}

\newcommand{\mH}{\ensuremath{m_{\PH}}\xspace}

\newcommand{\etg}{\ensuremath{\et^\gamma}\xspace}

\newcommand{\mT}{\ensuremath{m_{\mathrm{T}}^{\ell\ell\MET}}}
\newcommand{\MT}{\ensuremath{m_{\mathrm{T}}^{\gamma\MET}}}

\begin{document}
\begin{titlepage}
\pubblock

\vfill
\Title{Search for new physics in the low \MET monophoton channel with the CMS Detector}
\vfill
\Author{Toyoko Orimoto\support \\on behalf of the CMS Collaboration}
\Address{\napoli}
\vfill
\begin{Abstract}
With the recent discovery of the Higgs boson at the Large Hadron Collider (LHC), the goals of the Compact Muon Solenoid (CMS) experiment are now focused on probing for new physics beyond the standard model. The final state consisting of a low transverse energy photon and low missing transverse energy (\MET), also called the monophoton final state, can be used to constrain a variety of extensions of the standard model, including supersymmetry. I present a search for new physics in this low \MET monophoton channel using 7.3$\fbinv$ of 8 \TeV $pp$ collision data collected with the CMS detector. This analysis extends the high-energy single-photon searches to a lower energy regime. In the absence of deviations from the standard model predictions, limits are set on the production cross section of exotic decays of the Higgs boson. In addition, we present model independent limits for a generic signal in the monophoton final state.
\end{Abstract}
\vfill
\begin{Presented}
DPF 2015\\
The Meeting of the American Physical Society\\
Division of Particles and Fields\\
Ann Arbor, Michigan, August 4--8, 2015\\
\end{Presented}
\vfill
\end{titlepage}
\def\thefootnote{\fnsymbol{footnote}}
\setcounter{footnote}{0}

\section{Introduction}

With the recent discovery of the Higgs boson at the Large Hadron Collider (LHC)~\cite{CMSHiggs, ATLASHiggs, CMSHiggs2}, the Compact Muon Solenoid (CMS) Experiment~\cite{Chatrchyan:2008zzk} is now focusing on searches for new physics at the energy frontier. The final state consisting of a photon and missing transverse energy (\MET), also called the monophoton final state, can be used to constrain a variety of beyond-the-standard-model (BSM) scenarios~\cite{exoHiggs}. 

\begin{figure}[b]
\centering
\includegraphics[height=1.8in]{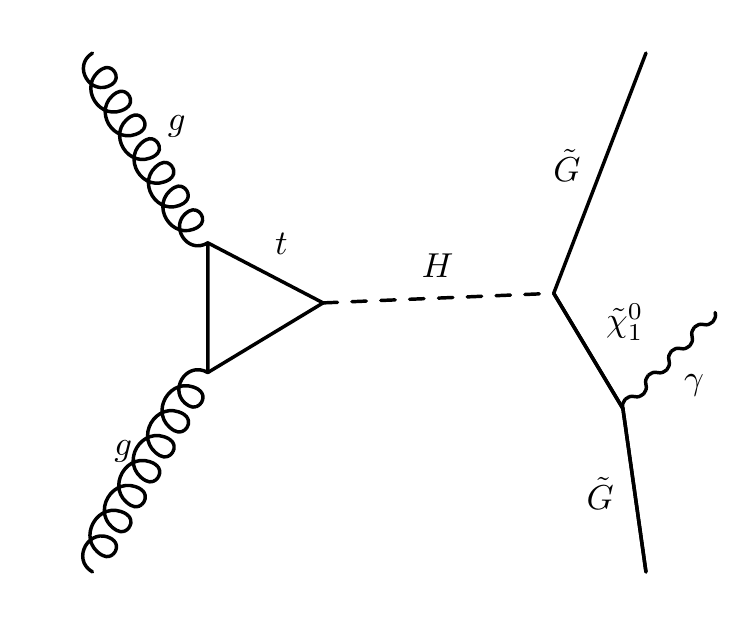}
\caption{Feynman diagram for the $\PH\to \gamma + \MET$  final state produced via $\Pg\Pg\PH$.}
\label{fig:feynman}
\end{figure}

In models of low-scale supersymmetry (SUSY) breaking, the Higgs boson can decay into a gravitino (\PXXSG) and a neutralino
($\PSGczDo$) or a pair of neutralinos~\cite{Djouadi:1997gw, Petersson:2012dp}. The neutralino then decays promptly into a photon and a gravitino, the lightest supersymmetric particle in this scenario, and a dark matter candidate.  The gravitino in this model has negligible mass. 
For neutralino masses $m_{\PSGczDo} < m_\PH/2$, with $m_\PH = 125\GeV$, the final state with  $\PH \to \PSGczDo\PSGczDo \to \gamma \gamma \PXXSG\PXXSG$ is expected to dominate.
On the other hand, for neutralino masses $m_{\PSGczDo} > m_\PH /2$, the decay into one neutralino ($\PH \to \PSGczDo \PXXSG\to \gamma \PXXSG\PXXSG$) is kinematically favored, resulting in the monophoton final state. Figure~\ref{fig:feynman} shows the Feynman diagram for this decay, with the Higgs boson ($\PH$) produced by gluon-gluon fusion ($\Pg\Pg\PH$).
Due to the 125 \GeV Higgs boson mass, the resulting photon transverse energy (\etg) and missing transverse energy (\MET) in the event will be relatively low compared to, for instance, high energy monophoton searches for dark matter~\cite{Khachatryan:2014rwa, Aad:2014tda}.  

The first search for decays of the Higgs boson to undetectable particles and one photon in the low energy regime is presented, using $7.4$\fbinv of data
 collected at $\sqrt{s} = 8\TeV$ with the CMS detector~\cite{CMS:2015ula, Khachatryan:2015vta}. The results are presented in a model-independent framework, as well as in the low-scale SUSY benchmark model. 


\section{The CMS detector}

The central feature of the CMS apparatus~\cite{Chatrchyan:2008zzk} is a superconducting solenoid of 6 m internal diameter, providing a magnetic field of 3.8 T. Within the superconducting solenoid volume are a silicon pixel and strip tracker, a lead tungstate crystal electromagnetic calorimeter (ECAL), and a brass and scintillator hadron calorimeter (HCAL), each composed of a barrel and two endcap sections. Forward calorimeters extend the pseudorapidity coverage provided by the barrel and endcap detectors. Muons are measured in gas-ionization detectors embedded in the steel flux-return yoke outside the solenoid.

\section{Data and simulation events}~\label{sec:MC}

At the trigger-level, the presence of one central ($\abs{\eta^{\gamma}}< 1.44$) photon candidate with $\etg > 30\GeV$ and passing loose isolation criteria is required. $\MET$ at trigger-level is calculated using calorimeter information only and is not corrected for muons. A requirement of $\MET  >$ 25\GeV is applied. The efficiency of this trigger with respect to $\etg$ and $\MET$ is measured using two control triggers. This trigger was made available through the CMS data parking program that was implemented for the last part of LHC Run 1 and allowed for an additional data stream with relaxed trigger requirements that was reconstructed during the long shutdown period. 

The signal and background processes (for $Z\gamma$, $\PW\gamma$, and $\gamma+$jets) were simulated using the \MADGRAPH event generator~\cite{Alwall:2014hca} at leading-order (LO). 
For $Z\gamma$ and $\PW\gamma$, next-to-leading order corrections to the cross section were derived with the \MCFM generator~\cite{Campbell:2010ff}.
The \PYTHIA 6.4 generator is used for parton showering and hadronization~\cite{Sjostrand:2006za}.
The CTEQ6L set of parton distribution functions (PDF)~\cite{Lai:2010nw} is used for LO
generators, while the CT10 PDF set~\cite{Lai:2010vv} is used for NLO generators.  
\GEANTfour ~\cite{Agostinelli:2002hh} is used to simulate the CMS detector response.
In  addition, the simulation samples are weighted to match the trigger efficiency and pileup distributions measured in data.

\section{Event selection}

The various backgrounds that contribute to this process are measured using either data-driven or simulation-based methods. Event selection criteria are applied to reject potential backgrounds in two different ways: first, a less stringent selection for a model-independent interpretation of the results and second, a  selection optimized for the supersymmetric Higgs decay into the monophoton final state described prior.

At preselection level, at least one well-isolated photon candidate with $\etg>45\GeV$ and $\abs{\eta^\gamma}<1.44$ is required, applying a cut-based photon identification selection~\cite{Khachatryan:2015iwa}.
To reduce SM backgrounds arising from the leptonic decays of $\PW$ and
$\cPZ$ bosons, a lepton veto is applied, rejecting events with $\ge 1$ electron (muon) with $\pt > 10\GeV$ and $\abs{\eta} < 2.5$ $(2.1)$ that is well-separated from the photon candidate ($\Delta R(\gamma,l) = \sqrt{\smash[b]{(\Delta\eta)^2 + (\Delta\phi)^2}} > 0.3$) and passes loose identification requirements~\cite{Khachatryan:2015hwa, Chatrchyan:2012xi}. In addition, the particle-flow $\MET$ in the event is required to be greater than $40\GeV$~\cite{Khachatryan:2014gga}. 

Beyond the preselection level described above, additional requirements are applied based on the model-independent selection or the optimized SUSY search strategy. For both approaches, jets are required to have $\pt^{\ensuremath{j}} > 30\GeV$ and $\abs{\eta^j} < 2.4$. These jets must not overlap with the photon candidate ($\Delta R(\gamma,\text{jet}) > 0.5$).

In the model-independent selection, events with two or more jets are rejected to discriminate against QCD multi-jet backgrounds.
For events with one jet, we require $\Delta \phi(\gamma,\text{jet})<2.5$ to reject $\gamma+$jet backgrounds, in which the photon and jet tend to be back-to-back in the transverse plane.

In the optimized selection for the SUSY benchmark analysis, a variety of quality requirements are applied to reject events with mismeasured $\MET$. 
The $\MET$ significance method computes an event-by-event estimate of the likelihood that the observed $\MET$
is consistent with zero, taking into account the reconstructed objects for each event and their
known resolutions~\cite{Khachatryan:2014gga}. The \MET significance in the event is required to be greater than 20.
%
A $\chi^2$ minimization method is also used~\cite{Khachatryan:2014rwa}, in which events without genuine $\MET$ have the mismeasured quantities re-distributed back into the particle momenta during the minimization process. 
Events are rejected if the minimized $\MET$ is less than $45\GeV$ and the $\chi^{2}$ probability is greater than $10^{-3}$.
In addition, the scalar sum of the transverse
momenta of the identified jets in the event ($\HT$) is required to be greater than 100\GeV. 
A requirement on the angle between the beam direction and the major axis of the supercluster~\cite{Khachatryan:2015iwa} is also applied to reject non-prompt photons with showers elongated along the beam line.
The transverse mass formed by the photon candidate, $\vmet$, and the angle between them ($\MT \equiv \sqrt{\smash[b]{2 \etg \MET [1-\cos\Delta\phi(\gamma,\MET)]}}$) is required to be greater than 100\GeV. 
To further reduce the $\Z\gamma$ background, which has a higher $\etg$ spectrum than the SUSY benchmark signal, 
a requirement of $\etg < 60\GeV$  is applied.

\section{Background estimation} \label{sec:backgrounds}

The $Z\gamma$, $W\gamma$, and $\gamma+$jets backgrounds are estimated using Monte Carlo (MC) samples, described in Sec~\ref{sec:MC}.
At the preselection level, the $\gamma+$jet process is the dominant background
due to the presence of an isolated photon and its large production cross section.
A data-driven technique is used to normalize the cross section of the $\gamma+$jets MC sample in two event classes. The correction factors of 1.7 for the $0$-jet bin and 1.1 for the $\ge 1$ jet bin were derived using a data control sample from a pre-scaled single photon control trigger with the \MET requirement reversed. A systematic uncertainty of $16\%$ is obtained for these correction factors based on the difference between the corrected and uncorrected MC samples and the relative fraction of 0-jet and $\ge 1$-jet events.
Scale factors (consistent with unity) are applied to all MC samples to correct for differences in efficiency modeling between data and simulation~\cite{Khachatryan:2015iwa}.

The remaining backgrounds, arising from jets and electrons which are misidentified as photons, are estimated with data control samples.
Jets can be misidentified as photons ($\text{jet} \to \gamma$) when an energetic $\pi^0$ in the jet decays to two collinear photons.
The  ($\text{jet} \to \gamma$) background is estimated in a data control sample of multijet events defined by reversing the \MET requirement ($\MET < 40\GeV$).
To estimate the contamination, the data control sample is used to measure the ratio of the number of candidates that pass the signal photon identification requirements to the number passing a looser set of photon identification requirements but failing one of the nominal isolation criteria. The numerator of this ratio is corrected for real photon contamination using template methods. The systematic uncertainty on this method is $35\%$, dominated by the choice of isolation sideband used for the real photon subtraction. 

Events with single electrons misidentified as photons ($\text{electron} \to \gamma$) are another significant background,  
largely arising from $\PW\to e\nu$ events.
The efficiency to identify electrons ($\epsilon_{\gamma_{\Pe}}$) is estimated using a tag-and-probe method with $\cPZ\to \Pep\Pem$ events in data~\cite{Khachatryan:2015hwa}. 
The inefficiency ($1-\epsilon_{\gamma_{\Pe}}$) of the selection to identify electrons is found to be $2.31\pm 0.03\%$. The systematic uncertainty of $6\%$ assigned to this measurement is dominated by the dependence of this inefficiency on the number of vertices reconstructed in the event and the number of tracks associated with the primary vertex.

Non-collision backgrounds, potentially arising from cosmic ray muons, beam halo muons and anomalous signals in the electromagnetic calorimeter, are estimated using template fits to the arrival time of the photon candidates. Such out-of-time backgrounds are found to be negligible in the signal region.

Two control regions are used to validate the accuracy of our background estimates. First we use a control region defined by the preselection level requirements. Second, we use a signal-free control region defined by the preselection level requirements, but with the lepton veto reversed. Both control regions confirm that the data are well modeled through the data-driven and simulation based background estimates.

\section{Results} \label{sec:results}

\begin{figure}[tbh]
\centering
\includegraphics[height=3.0in]{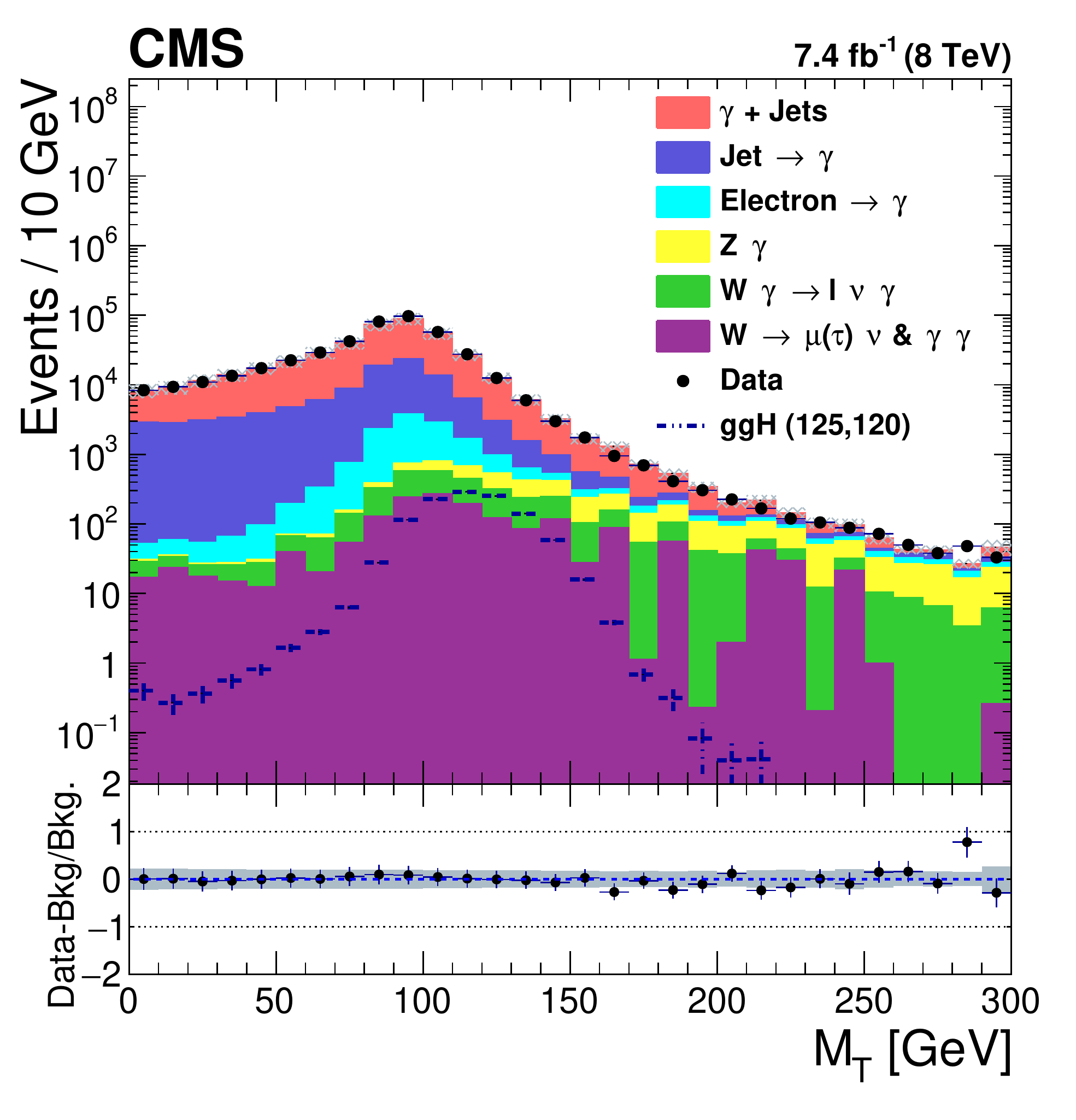}
\includegraphics[height=3.0in]{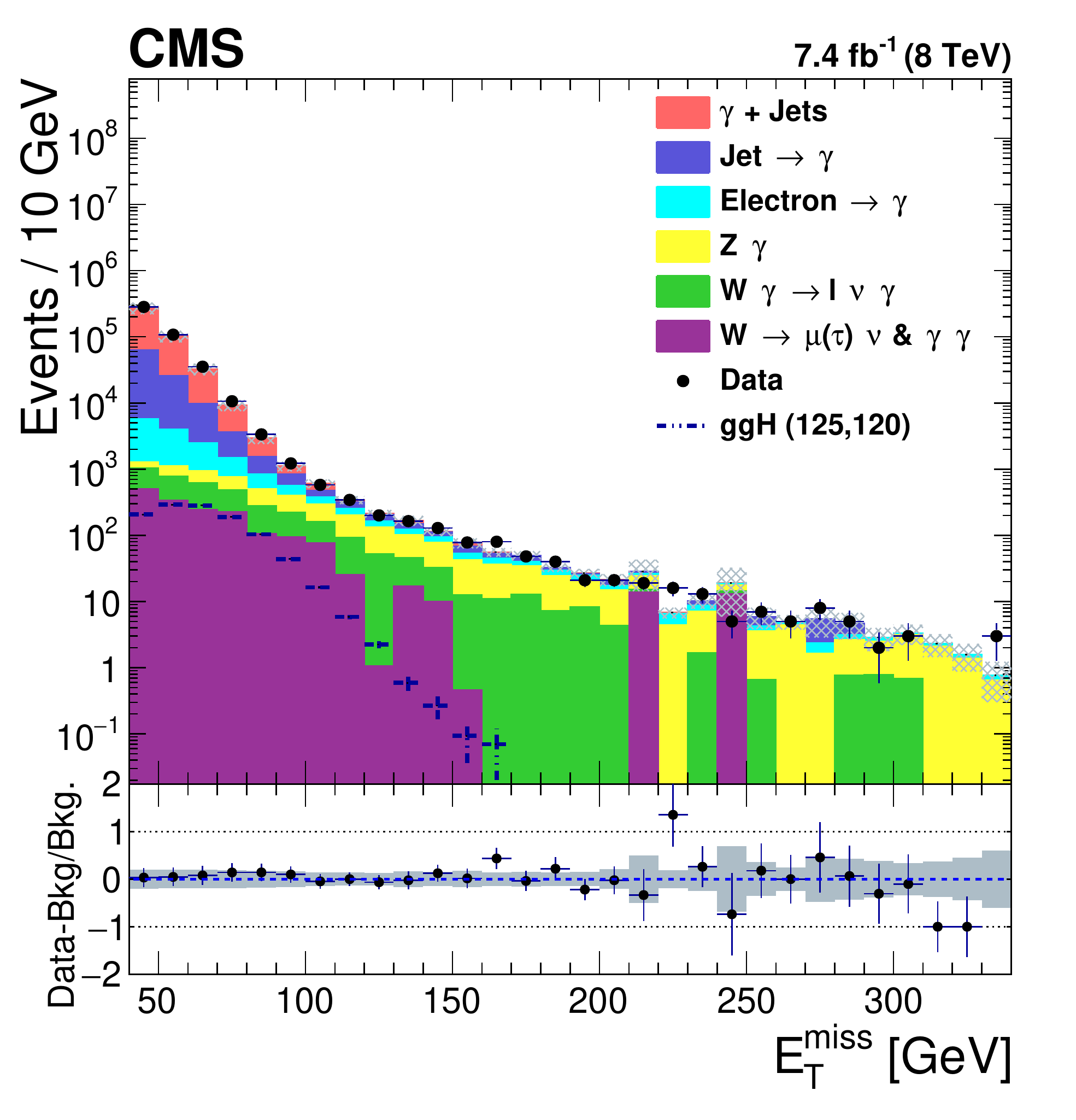}
\vspace{-0.5cm}
\caption{The $\MT$ (left) and $\MET$ (right) distributions for data, background
  estimates, and signal after the model-independent selection is applied.
The bottom panel in each plot shows the ratio of
  ($\text{data} - \text{background}$)/background and the gray band includes both the
  statistical and systematic uncertainties in the background
  prediction.
The signal is shown for $\mH = 125\GeV$ and $m_{\PSGczDo} =  120\GeV$.}
\label{fig:results}
\end{figure}

%
The total number of estimated background events and observed events in data are found to be compatible within uncertainties, as summarized in Table~\ref{table:yields}. 
Figure~\ref{fig:results} shows the $\mT$ and $\MET$ distributions for the model-independent selection.
Since no excess of events is observed, 95\% CL upper limits are calculated using an asymptotic CL$_S$ method~\cite{Read:2002hq}, in which the systematic uncertainties on the signal and background predictions are treated as nuisance parameters with log-normal prior distributions. The predominant systematic uncertainties on the background estimates arise from the ($\text{jet} \to \gamma$) measurement ($35\%$), the normalization of the $\gamma+$jet background ($16\%$), and the $\text{electron} \to \gamma$ measurement ($6\%$). The predominant systematic uncertainty on the signal SUSY model arises from the PDF uncertainty ($10\%$). 
A 2.6\% uncertainty on the measurement of the integrated luminosity is also applied~\cite{CMS:2013gfa}. 
The remaining subdominant systematic uncertainties are summarized in~\cite{Khachatryan:2015vta}.

\begin{figure}[tbh]
\centering
\includegraphics[height=2.2in]{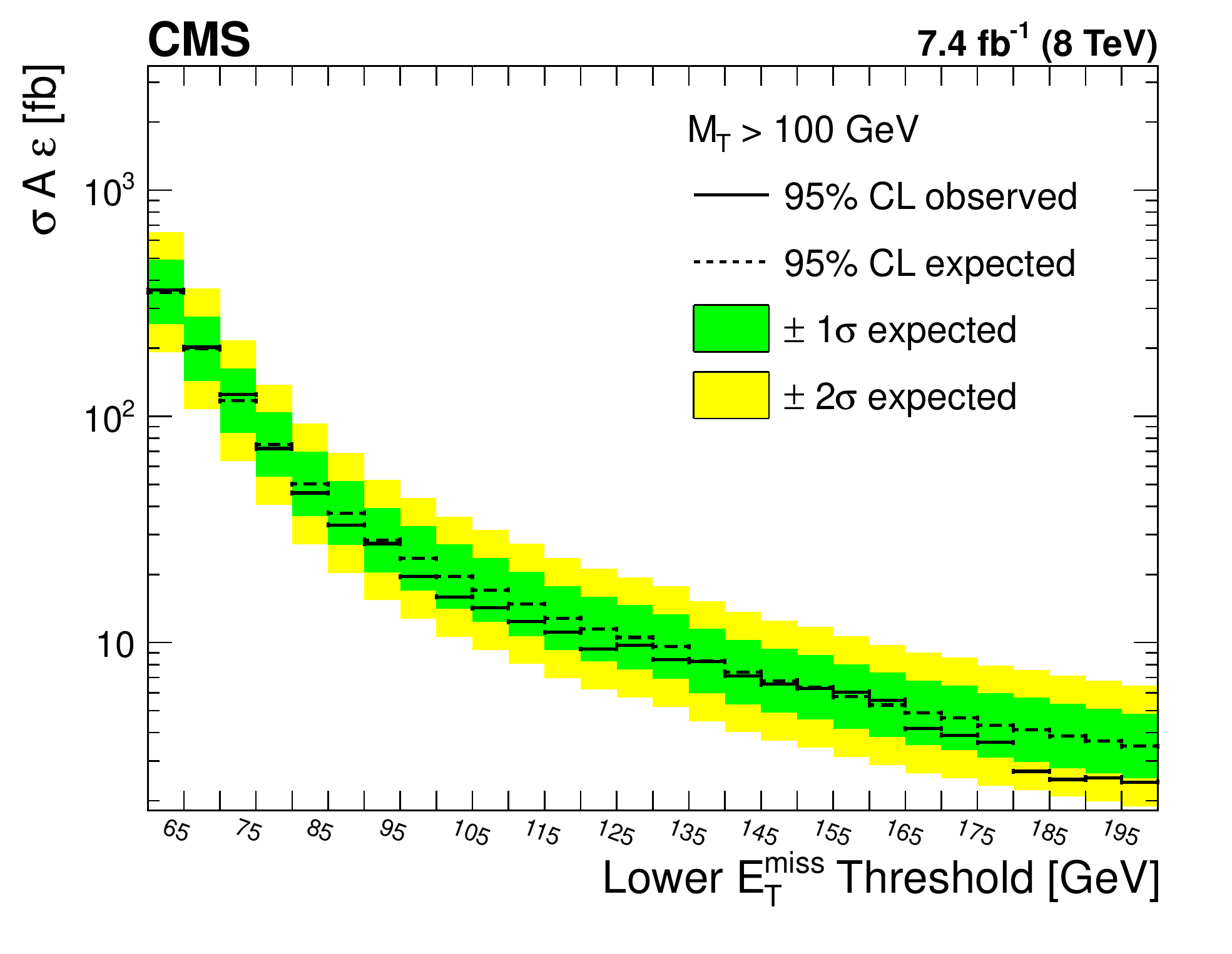}
\includegraphics[height=2.2in]{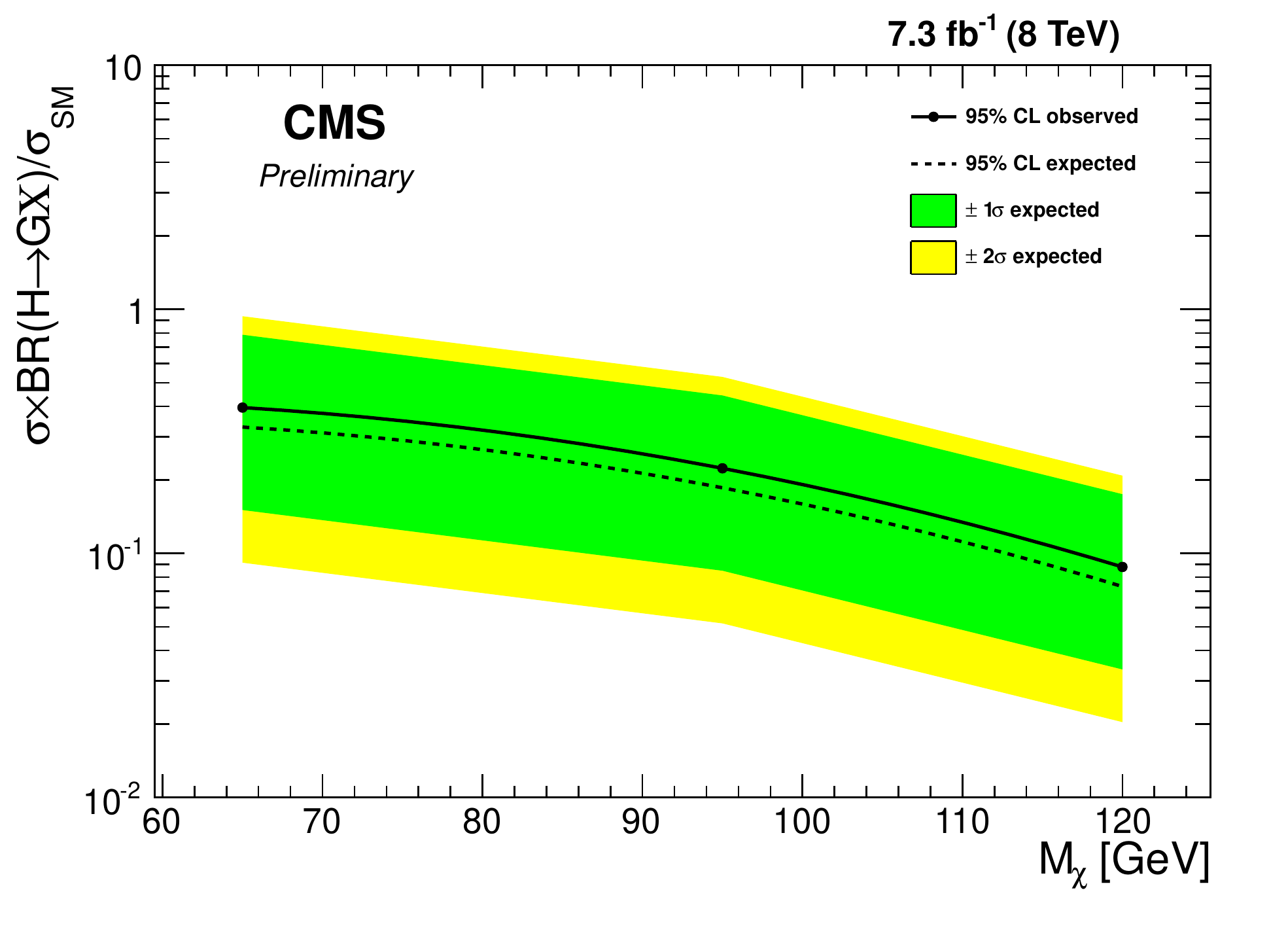}
\vspace{-0.5cm}
   \caption{
      (left) Expected and observed 95\% CL upper limit on 
the product of cross section, acceptance, and efficiency for a generic monophoton signal for $\MT> 100\GeV$, as function of
the $\MET$ threshold for the model-independent selection.
(right)
Expected and observed 95\% CL upper limits on  $\sigma \, \mathcal{B}/\sigma_{ \mathrm{SM}}$ for
     $\mH = 125\GeV$ as a function of $m_{\PSGczDo}$, assuming the SM Higgs boson cross section, for the SUSY analysis.   
}\label{fig:limits}
\end{figure}

In the model-independent case, 95\% CL upper limits on the cross section for producing a generic signal in the $\gamma+\MET$ final state are derived as a function of lower thresholds on $\mT$ and \MET. Figure~\ref{fig:limits} (left) shows the model-independent limits for $\mT > 100\GeV$. 
For the SUSY benchmark case, 95\% CL upper limits on ($\sigma \, \mathcal{B} )/\sigma_{SM}$, where $\sigma_{SM}$ is the cross
section for the SM Higgs boson, are evaluated for different mass values of $\PSGczDo$, as shown in Figure~\ref{fig:limits} (right).

\begin{table*}[tbhp]
 \centering
  \begin{tabular}{lcc}
\hline
\hline
Process & \multicolumn{2}{c}{Event yields}  \\
& Model-independent  & SUSY selection \\
\hline
\hline
$\gamma +$ jets                          & $(313 \pm 50 ) \times 10^3$ &  179 $\pm$  28 \\
$\text{jet}\to \gamma$            & $(910 \pm 320 ) \times 10^2$ &  269 $\pm$  94 \\
$\Pe \to \gamma$             & $10350 \pm 620$  &  355 $\pm$  28 \\
$\PW(\to \ell\nu)+\gamma $                &  $2239 \pm 111$ &  154 $\pm$  15 \\
$\cPZ( \to \nu \bar{\nu} )+\gamma$        &  $2050 \pm 102$ &  182 $\pm$  13 \\
Other                                 & $1809 \pm 91$ &   91 $\pm$  10 \\
\hline
Total background                       &   $(420 \pm 82 ) \times 10^3$    & 1232 $\pm$ 188 \\
\hline
Data                                   &  $442 \times 10^3$    & 1296  \\
\hline
$m_{\PSGczDo} =  65\GeV$ &  - & 653 $\pm$  77 \\
$m_{\PSGczDo} =  95\GeV$ & - & 1158 $\pm$ 137 \\
$m_{\PSGczDo} = 120\GeV$ &  - & 2935 $\pm$ 349 \\
\hline
\hline
 \end{tabular}
\caption{Observed yields and background estimates for the model-independent selection  (left) and for the SUSY benchmark selection  (right)  for $7.4\fbinv$ of 
$\sqrt{s} = 8\TeV$ $pp$ data. Signal predictions for different values of  $m_{\PSGczDo}$ are included for the SUSY analysis. These yields correspond to
 $\mathcal{B}(\PH \to \text{undetectable}+\gamma)=$100\%, assuming the SM
 cross section for producing the Higgs boson.
 The combination of statistical and systematic uncertainties is shown for the yields.}
\label{table:yields}
\end{table*}


\section{Conclusions}

In conclusion, first results for exotic decays of the Higgs boson to undetectable particles and one photon are presented with a dataset of $7.4\fbinv$ of 
$\sqrt{s} = 8\TeV$ $pp$ collision data collected with the CMS detector. The data are found to be compatible with background estimates, therefore, 95\% C.L. limits are presented in both model-independent and low-scale SUSY benchmark scenarios.



\newpage

\begin{thebibliography}{99}


\bibitem{CMSHiggs} 
  S.~Chatrchyan {\it et al.} [CMS Collaboration],
  Phys.\ Lett.\ B {\bf 716}, 30 (2012).

  \bibitem{ATLASHiggs} 
  G.~Aad {\it et al.} [ATLAS Collaboration],
  Phys.\ Lett.\ B {\bf 716}, 1 (2013).

\bibitem{CMSHiggs2} 
  S.~Chatrchyan {\it et al.} [CMS Collaboration],
  JHEP {\bf 1306}, 081 (2013).
   
\bibitem{Chatrchyan:2008zzk} 
  S.~Chatrchyan {\it et al.} [CMS Collaboration],
  JINST {\bf 3}, S08004 (2008).


\bibitem{exoHiggs} 
  D.~Curtin {\it et al.},
  Phys.\ Rev.\ D {\bf 90}, no. 7, 075004 (2014).
  
  
  \bibitem{Djouadi:1997gw} 
  A.~Djouadi and M.~Drees,
  Phys.\ Lett.\ B {\bf 407}, 243 (1997).
  
  
  \bibitem{Petersson:2012dp} 
  C.~Petersson, A.~Romagnoni and R.~Torre,
  JHEP {\bf 1210}, 016 (2012).
  
  
  
  \bibitem{Khachatryan:2014rwa} 
  V.~Khachatryan {\it et al.} [CMS Collaboration],
  arXiv:1410.8812 [hep-ex].
  
\bibitem{Aad:2014tda} 
  G.~Aad {\it et al.} [ATLAS Collaboration],
  Phys.\ Rev.\ D {\bf 91}, no. 1, 012008 (2015).
  
\bibitem{CMS:2015ula} 
  CMS Collaboration [CMS Collaboration],
  CMS-PAS-HIG-14-024.
  
\bibitem{Khachatryan:2015vta} 
  V.~Khachatryan {\it et al.} [CMS Collaboration],
  arXiv:1507.00359 [hep-ex].
  
  
  \bibitem{Alwall:2014hca} 
  J.~Alwall {\it et al.},
  JHEP {\bf 1407}, 079 (2014).
  
\bibitem{Campbell:2010ff} 
  J.~M.~Campbell and R.~K.~Ellis,
  Nucl.\ Phys.\ Proc.\ Suppl.\  {\bf 205-206}, 10 (2010)
  [arXiv:1007.3492 [hep-ph]].
  
  \bibitem{Sjostrand:2006za} 
  T.~Sjostrand, S.~Mrenna and P.~Z.~Skands,
  JHEP {\bf 0605}, 026 (2006).
  
  
  \bibitem{Lai:2010nw} 
  H.~L.~Lai, J.~Huston, Z.~Li, P.~Nadolsky, J.~Pumplin, D.~Stump and C.-P.~Yuan,
  Phys.\ Rev.\ D {\bf 82}, 054021 (2010).
  

  \bibitem{Lai:2010vv} 
  H.~L.~Lai, M.~Guzzi, J.~Huston, Z.~Li, P.~M.~Nadolsky, J.~Pumplin and C.-P.~Yuan,
  Phys.\ Rev.\ D {\bf 82}, 074024 (2010).
  
    \bibitem{Agostinelli:2002hh} 
  S.~Agostinelli {\it et al.} [GEANT4 Collaboration],
  Nucl.\ Instrum.\ Meth.\ A {\bf 506}, 250 (2003).
  
\bibitem{Khachatryan:2015iwa} 
  V.~Khachatryan {\it et al.} [CMS Collaboration],
  JINST {\bf 10}, no. 08, P08010 (2015).
  
\bibitem{Khachatryan:2015hwa} 
  V.~Khachatryan {\it et al.} [CMS Collaboration],
  JINST {\bf 10}, no. 06, P06005 (2015).
  
  \bibitem{Chatrchyan:2012xi} 
  S.~Chatrchyan {\it et al.} [CMS Collaboration],
  JINST {\bf 7}, P10002 (2012).
  
  
\bibitem{Khachatryan:2014gga} 
  V.~Khachatryan {\it et al.} [CMS Collaboration],
  JINST {\bf 10}, no. 02, P02006 (2015).
  
\bibitem{Read:2002hq} 
  A.~L.~Read,
  J.\ Phys.\ G {\bf 28}, 2693 (2002).
  

\bibitem{CMS:2013gfa} 
  CMS Collaboration [CMS Collaboration],
  CMS-PAS-LUM-13-001.



\end{thebibliography}
\end{document}